\documentclass[10pt,preprint]{aastex}

\usepackage{amsmath,natbib}

\begin{document}

\title{Magnetohydrodynamic Simulations of Shock Interactions with Radiative Clouds}

\author{P. Chris Fragile\altaffilmark{1}, 
Peter Anninos\altaffilmark{1}, 
Kyle Gustafson\altaffilmark{2}, and 
Stephen D. Murray\altaffilmark{1}}

\altaffiltext{1}{University of California,
Lawrence Livermore National Laboratory, Livermore, CA 94550}

\altaffiltext{2}{Department of Physics, University of Missouri - Columbia, Columbia, MO 65201}


\begin{abstract}
We present results from two-dimensional numerical simulations of the 
interactions between magnetized shocks and radiative clouds.  
Our primary goal is to characterize 
the dynamical evolution of the shocked clouds.  
We perform runs in both the strong and weak magnetic field limits and
consider three different field orientations.  
For the geometries considered, 
we generally find that magnetic fields external to, 
but concentrated near, the surface
of the cloud suppress the growth of destructive
hydrodynamic instabilities. External fields also 
increase the compression of the cloud by effectively acting as a 
confinement mechanism driven by the interstellar flow and
local field stretching. This can have a dramatic effect on both the
efficiency of radiative cooling, which tends to increase
with increasing magnetic field strength, and on the size and
distribution of condensed cooled fragments. In contrast, fields
acting predominately internally to the cloud tend to resist compression,
thereby inhibiting cooling.
We observe that, even at modest strengths ($\beta_0\lesssim100$), 
internal fields can completely suppress low-temperature 
($T<100$ K) cooling.
\end{abstract}


\section{Introduction}
\label{sec:introduction}

Shock waves are an important and common feature in both interstellar 
(ISM) and intergalactic (IGM) media.  They are 
triggered by such energetic phenomena as jets, supernovas, 
cloud-cloud collisions, and stellar winds and provide a means 
for transferring energy from such events into the ambient gas.
Since the ISM and IGM are generally inhomogeneous, an 
important problem in astrophysics is understanding 
the interaction of these shocks with overdense clumps or clouds.  
A thorough review 
of this problem in the unmagnetized and non-radiative limits is 
provided by \citet{kle94}.

Of special interest to us is large-scale 
shock-induced star formation, particularly 
in the neighborhoods of extragalactic radio jets.  
One of the first objects demonstrated to show a
correlation between a radio jet and regions of active star formation 
was the nearest radio galaxy, Centaurus A 
\citep[e.g.][]{bla75}.  Other examples have been found
as the sensitivity and spatial resolution of radio and optical
telescopes has improved.  For instance, ``Minkowski's Object'' is 
a peculiar small starburst system at the end of a radio jet emanating
from the elliptical galaxy NGC 541, located near the center of the
cluster of galaxies Abell~194 \citep{van85a}.  Correlations between radio 
and optical emissions have also been observed
in the so-called ``alignment effect'' in distant ($z>0.6$) radio galaxies
\citep{cha87,mcc87}.

These observations are most convincingly explained by
models in which shocks generated by the radio jet propagate through an
inhomogeneous medium and trigger gravitational collapse in relatively
overdense regions, leading to a burst of star formation 
\citep{beg89,dey89,ree89}.  
In a recent study \citep{fra04a}, we investigated a key component of 
such models - the radiative shock-induced collapse of intergalactic clouds.  
For moderate cloud densities ($\gtrsim 1$ cm$^{-3}$) and shock Mach numbers
($\lesssim 20$), we found that 
cooling processes can be highly efficient and result
in more than 50\% of the initial cloud mass cooling to below 100 K.  
The cold, dense fragments that form are presumably the precursors 
to active star-forming regions.

In the current work we 
consider the effects of dynamically important magnetic fields
in radiative shock-cloud collisions.  
Magnetic fields are known to be a pervasive element of the ISM and 
IGM and are often relevant in characterizing the local and global 
dynamical behaviors of these media.
In shock-cloud interactions, magnetic fields can act to 
suppress destructive hydrodynamic 
instabilities by providing additional tension 
at the interface between the cloud and the post-shock background 
gas \citep{nit81,mac94}.  
Magnetic fields can also limit the growth of disruptive 
vortices that form in the wake of the cloud, again primarily 
due to tension in the magnetic field lines as they are wound up within 
the vortices.  Strong external magnetic fields can also increase the 
compression of the shocked cloud material, due to the increased external
magnetic pressure.  This compression 
enhances the radiative efficiency of the cloud and allows 
additional cooling beyond that achievable without magnetic fields.  However, 
strong internal magnetic fields can resist compression, thereby 
inhibiting cooling of the cloud.  
In this paper we explore these competing effects 
and the general role of magnetic fields in radiative shock-cloud 
collisions.  
This paper also reports on the addition of magnetic fields to our 
astrophysical hydrodynamics code, Cosmos.  We proceed in 
\S\ref{sec:models} by describing the models considered in 
this work.  In \S\ref{sec:methods} we describe our implementation of radiation 
MHD.  Our results are presented in \S\ref{sec:results} and 
discussed further in \S\ref{sec:discussion}.

\section{Models}
\label{sec:models}
Models of shock-induced star formation are generally built upon 
the assumption of an inhomogeneous two-phase medium consisting 
of warm or cold dense clouds or clumps embedded in 
a hot, tenuous background medium.  
Following \citet{ree89}, \citet{beg89}, and \citet{mcc93}, we assume a 
background temperature $T_{b,i}=10^7$ K and density
$n_{b,i} = 0.01$ cm$^{-3}$.  In this work we restrict ourselves to 
modeling isolated clouds.  Each cloud has an initial temperature 
$T_{cl,i}=10^4$ K, appropriate for warm, ionized gas.  
Assuming pressure equilibrium 
between the two phases, the density of the cloud 
is $n_{cl,i}=(T_{b,i}/T_{cl,i}) n_{b,i}=\chi n_{b,i} = 10$ cm$^{-3}$, 
where $\chi=10^3$ is the density ratio between the cloud and background gases.
A planar shock of velocity 
$v_{sh,b} = 3.7 \times 10^3$ km s$^{-1}$ 
($\mathcal{M}=10$, where $\mathcal{M}$ is the Mach number,
measured in the background gas) is set up to propagate across this cloud.  
This background shock triggers a secondary, compressive shock of velocity 
\begin{equation}
v_{sh,cl} \simeq \left( \frac{n_{b,i}}{n_{cl,i}} \right)^{1/2} v_{sh,b} 
= \frac{v_{sh,b}}{\chi^{1/2}} ~.
\end{equation}
inside the cloud.

We require the clouds in these models to be 
large enough not to be destroyed by the shock prior to the 
onset of cooling.  This is equivalent to requiring $t_{cool}\ll t_{cc}$ 
or $q_s = \lambda_{cool}/R_{cl}\ll 1$, 
where  
\begin{equation}
t_{cool} \approx \frac{1.5 k_B T_{cl,ps}}{(n_{cl,ps}) \Lambda(T)} ~
\end{equation}
is the cooling time, 
\begin{equation}
t_{cc} = \frac{R_{cl}}{v_{sh,cl}} \simeq \chi^{1/2} \frac{R_{cl}}{v_{sh,b}} ~,
\label{eq:tcc}
\end{equation}
is the cloud-compression or cloud-destruction timescale, 
$q_s$ is the cooling parameter, and
$\lambda_{cool}\approx v_{sh,cl} t_{cool}$ is the cooling 
length.  
Provided $t_{cool}\ll t_{cc}$ (or $q_s\ll 1$), 
the shock is radiative and the post-shock gas loses 
much of its internal energy during compression.  
If radiative losses significantly outpace compressive 
heating, an initial compressive shock can 
trigger cooling to very low temperatures \citep{mel02,fra04a}.  
If, however, $t_{cool} > t_{cc}$ (or $q_s>1$), the shock-heated gas does 
not have time to cool before it is diffused into 
the background gas \citep{kle94}.  Approximating the cooling function as 
$\Lambda = 1.33 \times 10^{-19} T^{-1/2}$ erg cm$^3$ s$^{-1}$ 
\citep{kah76}, 
the restriction on the cloud radius becomes
\begin{equation}
\left( \frac{R_{cl}}{100 \mathrm{~pc}}\right) \gtrsim 1.2\times10^{-4} 
 	\left(\frac{\chi}{10^3}\right)^{-2}
	\left(\frac{v_{sh,b}}{10^3 \mathrm{~km~s}^{-1}}\right)^4
	\left(\frac{n_{cl,i}}{1 \mathrm{~cm}^{-3}}\right)^{-1} ~.
\end{equation}
For the parameters used in this work, $R_{cl}$ needs only to be larger 
than about 0.2 pc ($\lambda_{cool}\approx0.02$ pc).  
However, we choose $R_{cl} = 100$ pc in order to 
link our work with previous unmagnetized radiative shock-cloud 
simulations [run A in 
\citet{mel02} and model E3 in \citet{fra04a}].  
The larger cloud is also more reasonable for triggering a large 
burst of star-formation.  
All of the cooling runs in this work therefore begin with 
$q_s\approx 2\times10^{-4}$, $t_{cool}=190$ yr, and 
$t_{cc}=0.85$ Myr.

In this work we consider an array of simulations in the strong 
and weak magnetic field limits.  We include runs with initial field 
strengths of 
$\beta_0=1$, 4, 100, and $\infty$, where $\beta=P/(B^2/8\pi)$ 
is the ratio of hydrodynamic to magnetic pressure
in the pre-shock region.  
Although magnetic field strengths are difficult to measure in 
astrophysics, $\beta$ appears to be of order 10 in the diffuse 
regions of the ISM \citep{mac04}.  
A value of $\beta_0=4$ in our models corresponds to 
an initial field strength in the pre-shock region of 9.3 $\mu$G, 
comparable to inferred interstellar field strengths \citep{ran89,fit93}.
As in \citet{fra04a}, our models are two-dimensional, in Cartesian
geometry.  In this geometry, we consider
three different field orientations:  (1) parallel to both the 
planar shock front and the 
cylindrical cloud ($B_z$), (2) parallel to the shock front but 
perpendicular to the cloud ($B_y$), and (3) perpendicular to the 
shock front and the cloud ($B_x$).  
For the strong shocks considered here, $B_z$ and 
$B_y$ are enhanced by about a factor of 
$(\gamma+1)/(\gamma-1) = 4$ (for a $\gamma=5/3$ gas) in the post-shock 
region, whereas 
$B_x$ is continuous across the shock. 
We also consider runs with and without 
radiative cooling active for each of the field orientations.
Together, these runs facilitate an easy comparison of results 
with various magnetic field configurations and atomic
processes using a single numerical scheme.
The various runs and physical parameters of each
shock-cloud simulation considered in this study are 
summarized in Table \ref{tab:models}.

For the strong magnetic field cases we distinguish 
between what we consider
primarily {\em internal} fields ($B_z$ component) and what we consider
primarily {\em external} fields ($B_x$ and $B_y$ components), differentiated
by the regions where their effects are greatest.  We make this distinction
based upon the fact that 
the $B_z$ field component, being parallel to the cylindrical cloud, 
plays no role other than to modify the total effective pressure (there are no 
gradients in the $z$-direction, so the ${\bf B} \cdot \nabla$ 
terms in equations (\ref{eqn:mom}) and (\ref{eqn:b}) below drop out).  
However, for the strong shocks considered here, the post-shock gas pressure 
is generally much higher than the magnetic pressure,
($P/[B^2/8\pi] =\beta \approx {2\gamma[\gamma-1]^2/[\gamma+1]^3}\mathcal{M}^2 \beta_0 \gg 1$),
and so the $B_z$ component 
plays little role in the non-radiative case \citep{jon96}.  
For a strongly radiating shock, however, the thermal gas pressure 
can drop quite dramatically behind the shock.  
Magnetic pressure 
can then become dominant in the post-shock layer and 
prevent further compression 
once $B_{z,cl}^2/8\pi = \rho_{cl,i} v_{sh,cl}^2$ \citep{mck80} or 
\begin{equation}
\left(\frac{\rho_{max}}{\rho_{cl,i}} \right) = 
	\left(\frac{\mu m_H v_{sh,b}^2 \beta_0}{k_B T_{b,i}} \right)^{1/2} ~.
\label{eq:rhomax}
\end{equation}
Prior to this, the cooling will proceed approximately isobarically.  
After this, any further cooling 
may only proceed approximately isochorically.
For our usual parameters, with $\beta_0=4$, 
this predicts a peak density enhancement of 
$\rho_{max}/\rho_{cl,i}\approx 26$.  
The field amplification 
$B_{max}/B_i$ {\em inside} 
the cloud should be identical.  If we now assume the cloud compresses 
isothermally due to radiative cooling, then we can estimate 
$\beta$ inside the cloud since both the thermal gas 
pressure and magnetic field strength scale directly with density 
in this case.  
Thus, $\beta=P/(B^2/8\pi)\propto \rho^{-1}$ inside the cloud.  
The actual equation of state in our radiatively cooled runs is much softer 
than $P\propto \rho$, so this line of reasoning only implies an upper limit 
$\beta_{min} < (\rho_{cl,i}/\rho_{max})\beta_0$.  We emphasize that 
this analysis strictly applies only to our radiatively-cooled $B_z$ runs.

As the dominant role of the $B_z$ field is restricted to the interior 
of the cloud, we consider it an internal field.
On the other hand, we refer to $B_y$ and $B_x$ components as external fields.  
This does {\em not} mean that these field lines 
do not penetrate the cloud or have no role in the cloud interior.  
Rather it refers to the fact that their dominant roles 
are along the cloud surface or even external to the cloud.
In fact, for our $B_x$ cases, we do not expect much of a change from our
unmagnetized results because the shock 
proceeds mostly parallel to the field lines, preventing the Lorentz force 
from acting.  The magnetic fields, therefore, are not able to feed back 
strongly on the hydrodynamic evolution.  In our $B_y$ cases, 
on the other hand, the fields have a dominant role in the 
cloud evolution.  These fields become trapped at the nose of the 
cloud allowing the magnetic pressure and field tension to continually 
increase, at least 
until the cloud is accelerated to the velocity of the post-shock 
flow.  We can get a rough estimate of when this will 
occur by considering just the 
Lorentz acceleration term, which for the cloud is approximately 
$B_{y,i}^2/(4\pi \rho_{b,i} R_{cl})$.  Assuming a constant acceleration, 
the velocity of the cloud will match that of the post-shock background 
gas after 
\begin{equation}
t\approx \frac{3 \mu m_H v_{sh,b} R_{cl} \beta_0}{8 k_B T_{b,i}} ~,
\end{equation}
where we have used the shock-jump condition 
$v_{b,ps}=3v_{sh,b}/4$ for a $\gamma=5/3$ gas.  
Using our normal parameters and $\beta_0=4$, we expect it will take 
the cloud approximately 6.4 Myr to reach the velocity of the 
post-shock background gas or about $8t_{cc}$.

We should note that our classification of field geometries as {\em internal} 
or {\em external} 
is particular to the two-dimensional Cartesian 
grid used in this work.  For instance, if our cloud were a three-dimensional 
sphere, there would be no distinction between our 
$B_y$ and $B_z$ cases.  
Nevertheless, the contrast between internal and external 
field effects should remain valid even in three-dimensions.  
Realistically, magnetized shock-cloud collisions in nature will 
probably contain aspects of each of our idealized runs.

For runs in which 
radiative cooling is ignored, \citet{kle94} showed that the hydrodynamic 
shock-cloud problem is invariant under the scaling 
\begin{equation}
t \rightarrow t\mathcal{M}~, \quad v \rightarrow v/\mathcal{M}~, \quad P \rightarrow P/\mathcal{M}^2~,
\end{equation}
with distance, density, and pre-shock pressure left unchanged, 
provided $\mathcal{M} \gg 1$.  
\citet{mac94} showed that the non-radiative 
magnetohydrodynamic case obeys the same scaling relations provided 
\begin{equation}
\bf{B} \rightarrow \bf{B}/\mathcal{M}~.
\end{equation}
Therefore, our non-radiative results are representative of {\em all} such 
cases provided $\mathcal{M} \gg 1$.  
However, these scaling relations do not hold when radiative cooling 
is important.  For radiative clouds, the results depend sensitively 
on the initial parameters \citep{fra04a}.

\section{Numerical Methods}
\label{sec:methods}

We carry out our simulations using Cosmos, a massively parallel, 
multi-dimensional, multi-physics magnetohydrodynamic code for both Newtonian 
and relativistic flows developed at Lawrence Livermore National 
Laboratory.  The relativistic capabilities and
tests of Cosmos are discussed in \citet{ann03a}.  Tests of the Newtonian
hydrodynamics options and of the microphysics relevant to the
current work are presented in \citet{ann03b} 
and will not be discussed in detail here.
The new elements introduced in this paper are the magnetic fields
and their coupling to the fluid motion and state. 
Currently the magnetic fields are only implemented as part of the 
zone-centered and staggered-mesh artificial viscosity hydrodynamic 
schemes in Cosmos.  The results in this paper use the zone-centered 
scheme.  As this is the
first work to introduce magnetic fields into Cosmos, we discuss
briefly the dynamical equations, reconnection corrections,
divergence cleansing, and numerical tests.

The magneto-hydrodynamics (MHD) equations solved in Cosmos take the form:
\begin{eqnarray}
 \frac{\partial \rho}{\partial t} +\nabla \cdot ({\bf v} \rho)
        &=& 0,
        \label{eqn:dens} \\
 \frac{\partial (\rho {\bf v})}{\partial t} +\nabla \cdot ({\bf v} \rho {\bf v})
        &=& -\nabla \left(P + \frac{B^2}{8\pi} \right) \nonumber \\
        & & + \frac{1}{4\pi} ({\bf B} \cdot \nabla) {\bf B} 
            - \rho \nabla \phi , 
        \label{eqn:mom} \\
 \frac{\partial e}{\partial t} +\nabla \cdot ({\bf v} e)
        &=& -P \nabla \cdot {\bf v}
            + \frac{\eta}{4\pi} (\nabla\times{\bf B})^2 \nonumber \\
        & & + \Lambda(T,\rho) ,
        \label{eqn:en} \\
 \frac{\partial {\bf B}}{\partial t} +\nabla \cdot ({\bf v} {\bf B})
        &=& ({\bf B} \cdot \nabla) {\bf v} 
            - \nabla\times(\eta\nabla\times{\bf B}) \nonumber \\
        & & - \nabla \psi,
        \label{eqn:b}
\end{eqnarray}
where ${\bf v}$ is the fluid velocity, $e$ is the fluid internal energy density,
$\rho$ is the fluid density, 
$P$ is the fluid pressure, ${\bf B}$ is the magnetic field,
and $\phi$ is the gravitational potential obtained from Poisson's 
equation $\nabla^2 \phi = 4 \pi G \rho$.  In practice, we ignore self gravity 
for the two-dimensional results presented here 
but include it in the equations in anticipation of future 
three-dimensional simulations of radiative magnetized shock-cloud collisions.
The cooling function $\Lambda(T,\rho)$ 
is solved using the equilibrium 
cooling curve model described in previous work \citep{ann03b,fra04a}.
This form of the MHD equations is derived with the standard
assumptions relevant for many astrophysical problems:
the system is nonrelativistic and fully ionized, the displacement currents
in Maxwell's equations are neglected,
the net electric charge is small, and the characteristic
length scales are large compared to particle
gyroradii scales.

In equations (\ref{eqn:en}) and (\ref{eqn:b}), 
$\eta$ is the non-ideal resistivity coefficient, used here 
to correct for magnetic reconnection errors 
that can occur in numerical schemes that solve the internal 
(rather than total) energy equation \citep{mag97,sto01}.  
Artificial resistivity spreads 
current sheets out over a few zones to keep them resolved, and compensates
partially for the energy lost in unresolved reconnection flows.  
This procedure is similar to the treatment of shocks in 
artificial viscosity schemes.  We expect that 
such anomalous reconnection could be important for the flows 
considered here.  Most of the results in this paper use the 
\citet{sto01} form of $\eta$ 
\begin{equation}
\eta_{SP} = \frac{k_1 (\Delta x)^2}{\sqrt{4\pi\rho}} \vert \nabla \times {\bf B} \vert ~,
\label{eq:etasp}
\end{equation}
where $\Delta x$ is the grid spacing and $k_1$ is a dimensionless 
parameter used to adjust the strength of the artificial resistivity.  
By making the 
artificial resistivity proportional to $\nabla \times {\bf B}$, this form 
ensures that it has negligible effect in smooth regions of the flow, 
yet is large inside current sheets.  However, we find 
that the simpler \citet{nit01} form 
\begin{equation}
\eta_N=k_1 v_A \Delta x ~,
\label{eq:etan}
\end{equation}
where $v_A=B/\sqrt{4\pi\rho}$ is the Alfven wave velocity, yields better 
results for the class of Riemann problems discussed in \citet{fal02} 
(see Appendix A).  For comparison, we present results of 
magnetized shock-cloud interactions using both 
forms of $\eta$, as well as a run without artificial resistivity.

The scalar potential $\psi$ in equation (\ref{eqn:b}) 
is introduced as a divergence cleanser to maintain a divergence-free 
magnetic field ($\nabla \cdot {\bf B} = 0$).  Options are included 
in Cosmos to
solve any one of the following constraint equations for $\psi$ \citep{ded02}:
\begin{eqnarray}
\nabla^2 \psi &=& - \frac{\partial \nabla\cdot {\bf B}}{\partial t} 
               \approx  - \frac{\nabla\cdot {\bf B}}{\Delta t} , \\
\psi &=& - c_p^2 \nabla \cdot {\bf B} , \\
\frac{\partial \psi}{\partial t} &=& - \frac{c_h^2}{c_p^2} \psi 
                                     - c_h^2 \nabla \cdot {\bf B} ,
\label{eqn:psi}
\end{eqnarray}
which correspond, respectively, to elliptic, parabolic, and
mixed hyperbolic and parabolic constraints.
Here $c_p$ and $c_h$ are user-specified constants used to
regulate the filtering process and weight the relative 
significance of the hyperbolic and parabolic components.  
For all of the calculations presented in this paper, we use
the strictly parabolic form, 
which we find to be the most effective
and least costly method to preserve the divergence constraint.
However, we have confirmed that its use makes relatively little 
difference in the dynamical evolution of the shocked clouds presented in the 
main body of this work.
This is likely a result of adopting the nonconservative form
of the MHD equations in which the acceleration terms
proportional to $\nabla \cdot {\bf B}$ have been explicitly
eliminated from equations (\ref{eqn:dens}) - (\ref{eqn:b})
\citep{bra80}.

We have validated the newly added magnetic field equations 
using a standard set of single and multi-dimensional MHD tests
including: advection of a localized pulse of transverse magnetic field, 
propagation of circularly polarized Alfv\'en waves, 
propagation of sheared Alfv\'en waves, 
magnetosonic rarefaction waves,
multiple MHD Riemann problems, and an MHD
shock-cloud collision problem.  A brief summary of
these test results is presented in the appendix of this paper.

The calculations in this work 
are carried out on a fixed, two-dimensional Cartesian
($x$,$y$) grid, implying that the simulated clouds are
cylindrical rather than spherical.  
The computational grid is $8R_{cl} \times 8R_{cl}$ with the cloud 
initially located at the center of the grid.  This 
is slightly larger 
than the grid we used in our previous work \citep{fra04a}.  
The larger grid allows us to 
maintain the leading edge of the bow shock on the grid.  In tests we 
found about a 10\% difference in some of the measured cloud parameters 
when comparing the smaller and larger grid, due primarily to the 
front edge of the bow shock reaching the inflow boundary of the smaller 
grid.

We use a constant inflow 
boundary condition for the post-shock 
gas along the left-most edge of the grid.  
The top and bottom boundaries use flat (zero-gradient) 
boundary conditions.  The right boundary uses outflow ($v_x\ge0$) 
conditions.  For runs with zero magnetic field 
we employ a reflective boundary along the symmetry axis of the 
problem and only evolve half of the grid.
As in our previous work 
we use a localized diffusion filter at all of the 
outflow boundaries to minimize 
strong reflections.  We find that this technique 
does an adequate job of preventing unphysical feedback from the 
boundaries while maintaining the integrity of the interior solution.

Runs BY4C(L), BY4C(L1), and BY4C(L2) are carried out at 
a fixed spatial resolution of 
$\Delta x = \Delta y = 1$ pc.  All of the remaining runs have twice that 
resolution ($\Delta x = \Delta y = 0.5$ pc).  The higher resolution runs 
have 200 zones per cloud radius, a value well above the resolution
requirements suggested by \citet{kle94} for non-radiative clouds.  
However, in the presence of cooling, which leads to extreme 
compressions and steep density gradients,
the resolution requirement becomes more stringent.  
In general, we find that we are only able to 
reliably follow the fragmentation of the cloud for approximately 
one hydrodynamic cloud-compression timescale (equation \ref{eq:tcc}). 
Beyond this time, 
further compression of the cloud is prevented by numerical 
resolution rather than any physical mechanism.

\section{Results}
\label{sec:results}

Figure \ref{fig:density} shows density contour plots for 
runs A, C, BZ4A, BZ4C, BY4A, BY4C, BX4A, and BX4C at $t=t_{cc}$.  
Several conclusions are immediately obvious from this 
figure: As noted in previous studies \citep{mel02,fra04a}, 
radiative cooling can have a critical effect on the evolution of 
shocked clouds; strong magnetic fields can also be dynamically 
important to the growth of instabilities and to the compression 
of the cloud \citep{mac94,jon96,gre99,gre00}; 
depending upon its orientation, the magnetic field can 
either enhance or resist cloud compression, actions which strongly affect 
the cooling efficiency of the cloud.

In the following sections we present more detailed analysis of 
our results as follows: 
In \S \ref{sec:instabilities} we consider the role of hydrodynamic 
instabilities; in \S \ref{sec:amplification} we discuss the evolution of 
the magnetic fields, particularly their amplification; in 
\S \ref{sec:compression} we quantify the compression of each model cloud, 
which is important to our discussion of cooling efficiency in 
\S \ref{sec:cooling}.  In \S\S \ref{sec:strength} and 
\ref{sec:resolution}, 
we explore the role of initial field strength and 
numerical resolution, respectively, on our results.  
Finally, in \S \ref{sec:resistivity}, we discuss reconnection and the 
role of artificial resistivity.

\subsection{Hydrodynamic Instabilities}
\label{sec:instabilities}
\begin{figure}
\plotone{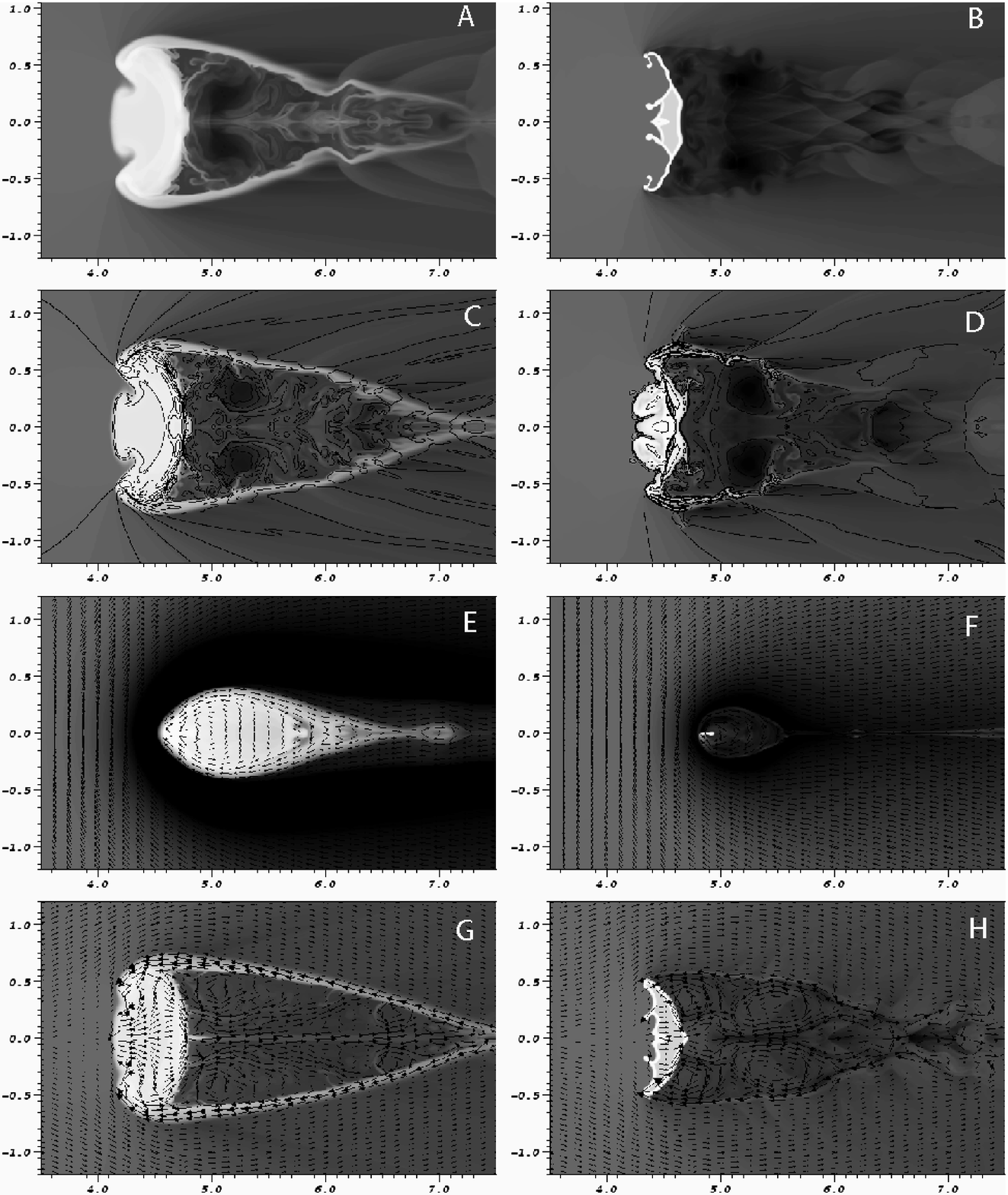}
\caption{Grayscale contour plots of $\log (\rho)$ for 
runs A (panel A), C (panel B), BZ4A (panel C), BZ4C (panel D), 
BY4A (panel E), BY4C (panel F), BX4A (panel G) and BX4C (panel H) 
at time $t=t_{cc}$.  For runs BZ4A and BZ4C we include 
contours of $\log (B^2/8\pi)$.  For runs BY4A, BY4C, BX4A, and 
BX4C we include a 
sampling of logarithmically-scaled arrows representing the local 
magnetic field; these arrows are scaled separately for each figure.
}
\label{fig:density}
\end{figure}

Previous studies \citep{kle94} have shown that strong shocks 
destroy unmagnetized, non-radiative clouds on a few dynamical timescales 
primarily through the growth of hydrodynamic (Kelvin-Helmholtz 
and Rayleigh-Taylor) instabilities.  The early growth of these 
instabilities is clearly seen in Figure \ref{fig:density}, 
particularly for run A 
(our non-radiative, unmagnetized case).  
These instabilities are seeded by the computational grid; hence, their 
nonlinear evolution is sensitive to 
the exact details of the simulation, including resolution and 
hydrodynamic method.  For instance, we noticed differences in the 
precise structure of the clouds at late times when we compared 
staggered-mesh and zone-centered versions of our code at the same 
resolution.  However, the important global characteristics of each simulation, 
such as field 
amplification, cloud compression, and cooling efficiency, are 
much less sensitive to these computational issues and we feel can 
therefore be reliably compared.

If the radiative efficiency of the gas has a sufficiently
shallow dependence upon the temperature, then radiative emissions 
cool the gas rapidly, in a runaway process, producing 
even higher densities as the cooling gas attempts to re-attain 
pressure balance with the surrounding medium \citep{fie65, mur89}.  
We find this can lead to an increase in the density contrast 
between the cloud and background of order $\gtrsim10^3$ above that
achieved in non-radiative cases, thus reducing the growth rate
of the Kelvin-Helmholtz instability 
\citep[$t_{KH}^{-1} = kv_{rel}/\chi^{1/2}$, ][]{cha61} 
by a factor $\gtrsim30$.
The slower growth rate helps stabilize the cloud as seen, 
for instance, by comparing the results 
of run A (non-radiative, unmagnetized case) and 
C (radiative, unmagnetized case) in Figure \ref{fig:density}.

\citet{mac94} and \citet{jon96} have shown numerically 
that predominantly external magnetic fields 
(combinations of $B_x$ and $B_y$) 
are very efficient at suppressing hydrodynamic instabilities, 
primarily due to tension in the magnetic field lines maintaining 
a more laminar flow around the cloud surface.  
Linear theory \citep{cha61} predicts Kelvin-Helmholtz 
instabilities will be suppressed if the local 
Alfv\'en speed exceeds roughly the velocity difference across the 
boundary, or $\beta < 2/\mathcal{M}^2$ for a $\gamma=5/3$ gas.  
The Rayleigh-Taylor instability will be suppressed if 
the Alfv\'en crossing time is less than the acceleration timescale, 
or $\beta < (2/\gamma)(\chi/\mathcal{M})^2$.  
Thus, for the parameters chosen in this work ($\chi = 10^3$ and 
$\mathcal{M} = 10$), Rayleigh-Taylor growth is strongly suppressed 
in all the magnetized runs considered ($\beta < 10^4$), 
while Kelvin-Helmholtz is only suppressed in runs which
evolve to a very strong field amplification ($\beta < 0.02$).
These conclusions are consistent with the results
observed in runs BY4A (non-radiative, $B_y$ case) and 
BX4A (non-radiative, $B_x$ case) compared to run A (non-radiative, 
unmagnetized case) in Figure \ref{fig:density}.  

The combined effect of strong external magnetic fields 
with cooling is to further 
suppress these instabilities.  This is illustrated by runs BY4C and 
BX4C in Figure \ref{fig:density}.

\subsection{Field Amplification}
\label{sec:amplification}
Magnetic fields can generally be amplified in one of two ways:  
(1) squeezing of field lines through compression, or 
(2) stretching of field lines through sheared motion.  
For external fields, stretching is much more 
important than squeezing \citep{mac94,jon96}, whereas 
for internal fields, compression inside 
the cloud provides the greatest amplification, 
particularly for radiative clouds.

\begin{figure}
\plotone{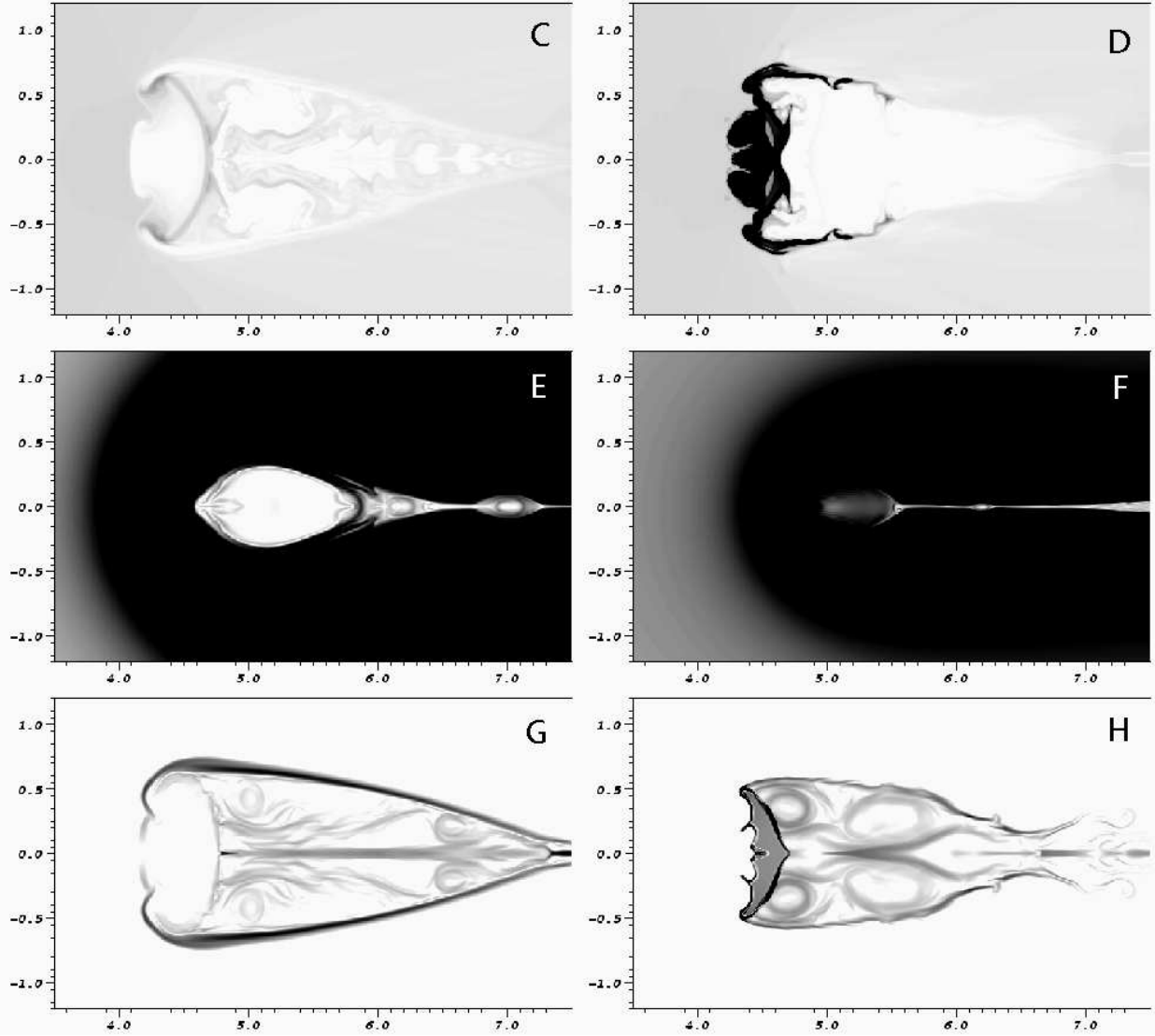}
\caption{Grayscale contour plots of $\log (\beta)$ for 
runs BZ4A (panel C), BZ4C (panel D), BY4A (panel E), BY4C (panel F), 
BX4A (panel G), and BX4C (panel H) at time $t=t_{cc}$.  
The scale of the plot goes from 
$\beta = 0.1$ ({\it black}) to $\beta=100$ ({\it white}).
}
\label{fig:beta}
\end{figure}

In Figure \ref{fig:beta} we present grayscale contour plots 
of $\log (\beta)$ for runs BZ4A, BZ4C, BY4A, BY4C, BX4A, and BX4C.  
Also, in Table \ref{tab:results} 
we record the minimum value of $\beta$ 
and the peak magnetic field enhancement ($B_{max}/B_i$) 
achieved in all simulations at $t=t_{cc}$.

For internal fields parallel to the cylindrical cloud ($B_z$), 
changes in the field strength simply follow changes in the density, 
as the fields are locked within the gas.  Thus, the location of the 
greatest magnetic field amplification coincides with the location of 
peak density amplification, generally near colliding shocks inside the 
cloud.  We therefore expect $B_{max}/B_i \ge \rho_{max}/\rho_{cl,i}$ 
for the $B_z$ runs, with $B_{max}/B_i > \rho_{max}/\rho_{cl,i}$ 
only if the peak field amplification occurs in the background gas.  
For non-radiative clouds the high density regions 
are also regions of high gas pressure, 
so the magnetic pressure fails to dominate anywhere inside the cloud 
(see panel C of Figure \ref{fig:beta}).  
However, for radiative 
clouds, the highest density regions cool most efficiently and have 
thermal gas pressures significantly below the magnetic pressure.  The 
magnetic fields thus provide an extra stiffness to such clouds relative 
to unmagnetized ones.  Notice the dramatically smaller $\beta_{min}$ 
in run BZ4C ($\beta_{min}=1.9 \times 10^{-3}$) 
compared to BZ4A ($\beta_{min}=4.1$), 
despite the fact that the peak magnetic field enhancement is 
somewhat comparable ($B_{max}/B_i=73$ for run BZ4C and 13 for run BZ4A).

For external fields perpendicular to the direction of shock propagation 
($B_y$), the greatest field amplification is at the front of the 
cloud.  This is because the background flow continues to stretch 
field lines around the nose of the cloud.  Since the clouds 
simulated here represent infinite 
cylinders, the field lines cannot ``slip'' around them as they 
might for a spherical cloud.  The cloud is enveloped in an 
ever-thickening cocoon of magnetic field lines.  
Thus, even an initially small field 
can build up to become dynamically important \citep{jon96}.  
However, because most of the field amplification is external to 
the cloud, where radiative cooling remains inefficient, $\beta_{min}$ 
is not significantly different between the non-radiative (BY4A) and 
radiative (BY4C) runs ($1.6\times10^{-5}$ and $3.8\times10^{-5}$, 
respectively).  The difference lies in the degree of cloud 
compression in the two runs due to runaway cooling in the radiative 
cloud.  We will return to this point below.

The increased tension in the field 
lines also provides an extra acceleration force 
on the cloud.  This explains the greater displacement of the clouds 
in runs BY4A and BY4C (relative to, say, runs BZ4A and BZ4C respectively)
in Figure \ref{fig:density}.  
In both cases, the cloud is accelerated to a velocity 
$v_{cl}\approx 120$ km s$^{-1}$ at the end of the simulation.  
Amplification of the field will continue until the cloud 
is accelerated to a velocity matching the post-shock flow 
$v_{b,ps}\approx 2800$ km s$^{-1}$.

For external fields parallel to the direction of shock propagation 
($B_x$), the field lines initially anchored in the cloud 
play the biggest role in its evolution.  The minimum value of $\beta$ 
initially occurs near the symmetry 
plane downwind of the cloud, where a ``flux tube'' forms \citep{mac94}.  This 
field enhancement is triggered by the rapid evacuation of gas from 
this region as a Mach stem forms in the shadow of the cloud.  
The field is also amplified by field stretching along the 
surface of the cloud and in vortices, primarily in the wake of the 
cloud.  However, both of these forms of amplification result 
in oppositely directed fields becoming adjacent.  
This configuration is unstable 
to reconnection, so the net amplification is limited.  Again, due to 
strong cooling in run BX4C, the thermal pressure inside the cloud 
drops significantly and allows the magnetic pressure 
to build up to a dynamically important value as evident in panel H 
of Figure \ref{fig:beta}.

\subsection{Cloud Compression}
\label{sec:compression}
One of our goals in this work is to quantify the efficiency
of cloud compression for each of the runs.
Previous authors \citep[e.g.][]{jon96} have tracked 
only the lateral expansion of the cloud. However,
such analysis does not fully account for longitudinal effects.  
Here we define cloud compression as $\zeta=A_{cl}(t)/A_{cl}^0$, 
where $A_{cl}^0$ is the initial cross-sectional area 
of cloud material 
and $A_{cl}(t)$ is the subsequent cross-sectional area at time $t$.  
In order to track the two gas components (cloud and background), 
we use two tracer fluids ($\mathcal{T}_{cl}$ and 
$\mathcal{T}_b$) which are passively advected
in the same manner as the density.  Throughout each calculation the 
distribution of $\mathcal{T}_{cl}$ reflects the distribution of original cloud 
material.  Numerically, the cloud compression is calculated as 
\begin{equation}
\zeta = \frac{\sum_{i,j} \mathcal{T}_{cl}(i,j;t)/[\mathcal{T}_{cl}(i,j;t) + \mathcal{T}_b(i,j;t)] \Delta x_i \Delta y_j }
             {\sum_{i,j} \mathcal{T}_{cl}^0(i,j) \Delta x_i \Delta y_j } ~,
\end{equation}
where $\mathcal{T}_{cl}/(\mathcal{T}_{cl}+\mathcal{T}_b)$ 
gives an estimate of the volume 
fraction of cloud material in a given cell and 
$\mathcal{T}_{cl}^0(i,j)=1$ inside the initial cloud and zero elsewhere.

\begin{figure}
\plottwo{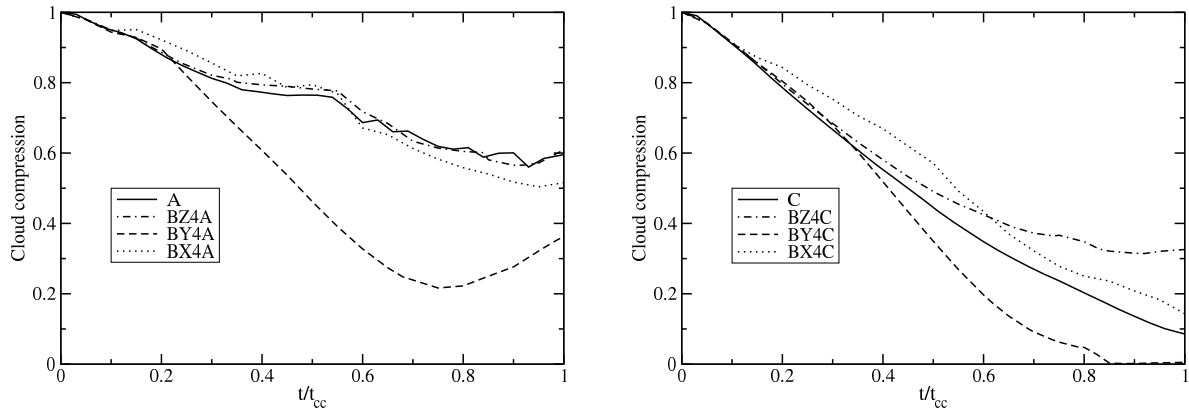}{f3b.eps}
\caption{Plot of cloud compression ($\zeta$) as a function of time for 
({\it a}) the non-radiative runs (A, BZ4A, BY4A, and BX4A) and 
({\it b}) the radiatively-cooled runs (C, BZ4C, BY4C, and BX4C).
}
\label{fig:compression}
\end{figure}

Figure~\ref{fig:compression} shows the cloud compression as a function of time 
for the non-radiative (A, BZ4A, BY4A, and BX4A) and the radiatively-cooled 
(C, BZ4C, BY4C, and BX4C) $\beta_0=4$ runs.  In Table \ref{tab:results} we 
record the peak density enhancement ($\rho_{max}/\rho_{cl,i}$) 
at $t=t_{cc}$ for all runs.  
All of the clouds are initially compressed over a 
timescale $t\approx t_{cc}$.  However, after the period 
of initial compression, the non-radiative clouds re-expand.  
This re-expansion phase leads to 
the cloud destruction phase as the growth of 
hydrodynamic instabilities is accelerated \citep{kle94}.
However, re-expansion is generally suppressed for radiative clouds.  In our 
simulations, the cross-sectional areas 
of these clouds continue to decrease until they reach 
a limit set by numerical 
resolution rather than any physical mechanism.  
The radiatively-cooled $B_z$ case (run BZ4C) is an exception.  
Here the internal magnetic field resists compression 
(thus inhibiting cooling) 
and allows the cloud to re-expand, similar to the non-radiative cases.  
Again, this makes the clouds more susceptible to destructive 
hydrodynamic instabilities as apparent in Figure \ref{fig:density}.  
As expected, we find that the peak density in each of the $B_z$ runs 
is about equal to the value set by equation (\ref{eq:rhomax}).

For runs BY4A and BY4C, the external field lines become trapped at the front 
of the cloud.  Because the clouds 
simulated here represent infinite 
cylinders, the field lines cannot ``slip'' around them as they 
might for a spherical cloud.  As more field lines wrap 
around the cloud, the compression becomes stronger.  The only 
direction the cloud is able to expand is in the direction of the 
original shock propagation, as occurs for run BY4A.  The extra compression 
in run BY4C causes runaway cooling at the highest rates we have observed, 
and the diminishing cloud quickly reaches the limits of our resolution.

In runs BX4A and BX4C, the magnetic field lines play little role in 
governing the compression of the clouds.  
Compression in these runs proceeds very 
similarly to the equivalent unmagnetized runs (A and C, respectively).

\subsection{Cooling Efficiency}
\label{sec:cooling}
As was shown by \citet{mel02} and \citet{fra04a}, the evolution of 
cooling-dominated clouds is quite different than that of 
non-radiative clouds.  Rather than 
re-expanding and quickly diffusing into the background gas,
the compressed cloud fragments into numerous dense, cold, compact filaments.  
These filaments survive for many 
dynamical timescales and presumably may be the precursors 
to active star-forming regions.  Contrast, for instance, the results 
of run A (non-radiative, unmagnetized case) and 
C (radiative, unmagnetized case) 
in Figure \ref{fig:density} to see the importance of radiative cooling.

\begin{figure}
\plottwo{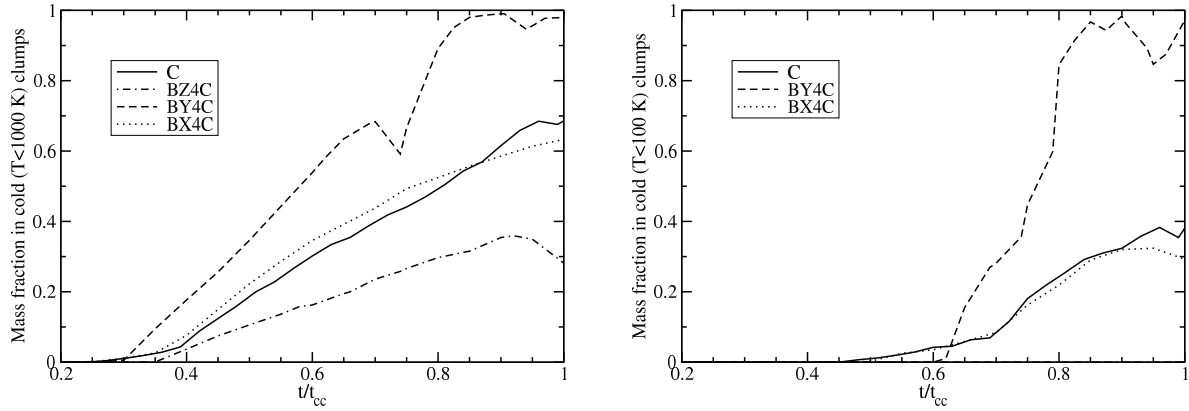}{f4b.eps}
\caption{Fraction of initial cloud material that has cooled below 
({\it a}) $T=1000$ K and ({\it b}) $T=100$ K as a function of time for runs 
C, BZ4C, BY4C, and BX4C.  Note that in run BZ4C none of the cloud material 
cools below $T=100$ K.
}
\label{fig:cool}
\end{figure}

Here we attempt to quantify the efficiency
of the cooling processes in runs C, BZ4C, BY4C, and BX4C, 
each of which included radiative cooling.
The same tracer fluid $\mathcal{T}_{cl}$ used to track the 
compression above also 
allows us to quantify how much of the 
initial cloud material cools below certain threshold temperatures.  
Figure~\ref{fig:cool} shows the fraction of cloud material that cools below 
$T = 1000$ and $T=100$ K as a function of time.  
The percentage of gas that cools to below 1000~K gives a
strong upper limit to the percentage that might form stars, while,
when well resolved, the amount that cools to below 100~K gives a more accurate 
measure.  In Table \ref{tab:results} we also record the minimum temperature 
achieved in each model at $t=t_{cc}$.

We note from the results that the cooling process is generally
extremely efficient throughout the cloud, although some differences 
are noted for the different field configurations.  
Since cooling efficiency is driven predominately by local gas density, runs 
such as BZ4C, in which the internal magnetic field lines stiffen the cloud 
and reduce the compression, cooling is not as efficient as in the 
fiducial unmagnetized run (C).  At the other extreme, strong external 
magnetic fields, such as those in run BY4C, 
greatly enhance the cooling efficiency, triggering greater rates of
runaway cooling.  Finally, in run BX4C, since the field 
plays little role in enhancing or reducing compression, it also has 
little effect on the overall cooling efficiency.

\subsection{Role of Initial Field Strength}
\label{sec:strength}
Thus far the magnetic field discussion has focused primarily on 
the effects of field orientation.  Now we explore the effects of 
varying the initial field strength.  Our general $\beta_0=4$ case 
corresponds to a reasonable initial field strength in the 
pre-shock region of 9.3 $\mu$G, comparable
to inferred interstellar and intergalactic field strengths 
\citep{ran89,fit93}.  
Nevertheless, astrophysical magnetic field strengths can vary by 
many orders of magnitude \citep{val03}.  
In the neighborhood of radio jets, the field strengths may be at 
least an order of magnitude higher than what we have chosen 
\citep[e.g.][]{kra04}.  
Furthermore, the dynamical importance of the magnetic fields depends 
on the thermal gas pressure, which itself can vary by 
orders of magnitude.  This suggests a very wide range of values of 
$\beta$ are plausible.  
Limits on computational resources 
prevent us from presenting a complete 
parameter study of field strengths, but we can nevertheless 
identify some important 
limits with a reasonably small set of simulations.

\begin{figure}
\plottwo{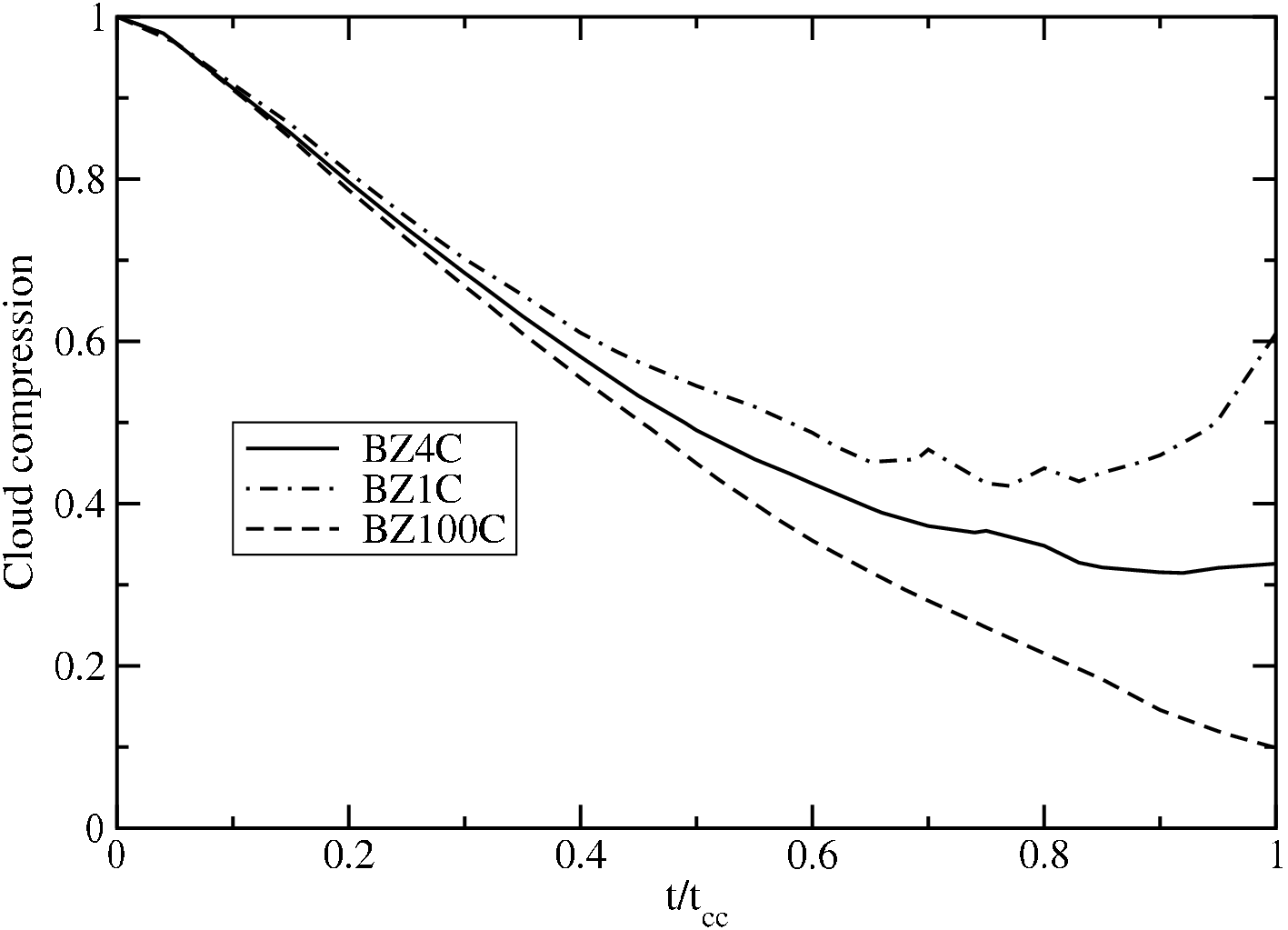}{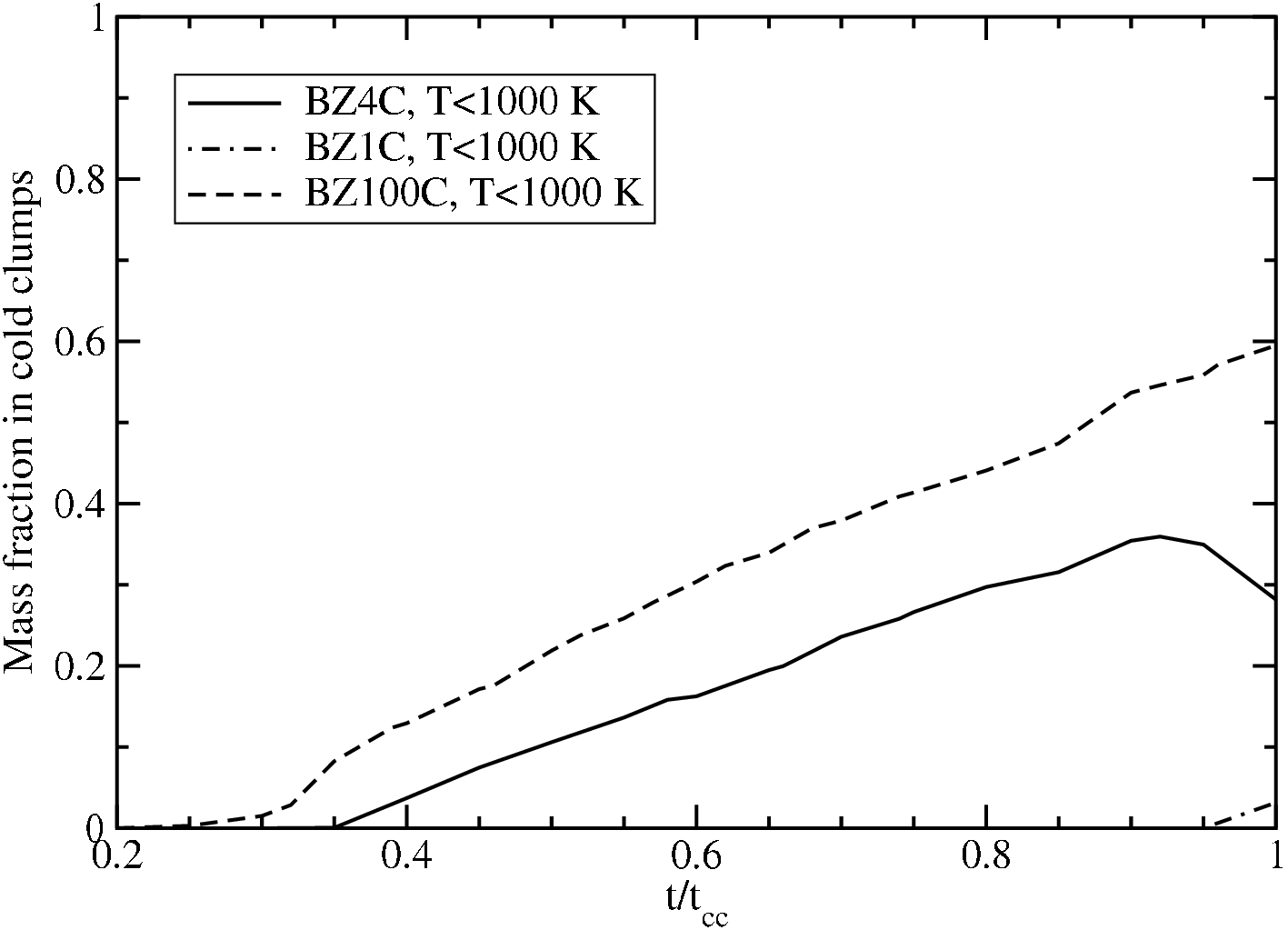}
\caption{Comparison of cloud compression ({\it a}) and cooling 
efficiency ({\it b}) for the radiatively-cooled $B_z$ runs.
}
\label{fig:bz}
\end{figure}

For the internal field ($B_z$) case (BZ4C) with $\beta_0=4$, 
the added stiffness 
of the magnetic pressure prevents any cloud gas from cooling 
below 100 K.  One can then ask, how strong must the initial field 
be in order to prevent cooling below our higher 
temperature threshold of 1000 K?  
Conversely one can ask, how weak does the initial field need to be 
in order not to significantly inhibit cooling over the timescales 
considered?  To attempt to answer these questions, we consider two 
additional field strengths for $B_z$: $\beta_0=1$ (run BZ1C) and 
$\beta_0=100$ (run BZ100C).  
In Figure \ref{fig:bz}, we compare the cloud compression and 
cooling efficiencies for the various $B_z$ runs.  
We see that for the $\beta_0=1$ case, low temperature cooling is 
almost completely suppressed; only a small amount of gas is able to 
cool below 1000 K.  The cloud also begins to re-expand towards the 
end of the simulation, a behavior commonly seen in simulations of 
non-radiative clouds.  
For $\beta_0=100$, the cloud behaves similarly to 
an unmagnetized, radiative cloud, although we note that 
there is still no cooling below 
100 K.  However, as noted in our previous work \citep{fra04a}, 
this particular diagnostic is very sensitive to small changes in the 
simulations, especially spatial resolution,
so its usefulness is somewhat limited.

\begin{figure}
\plottwo{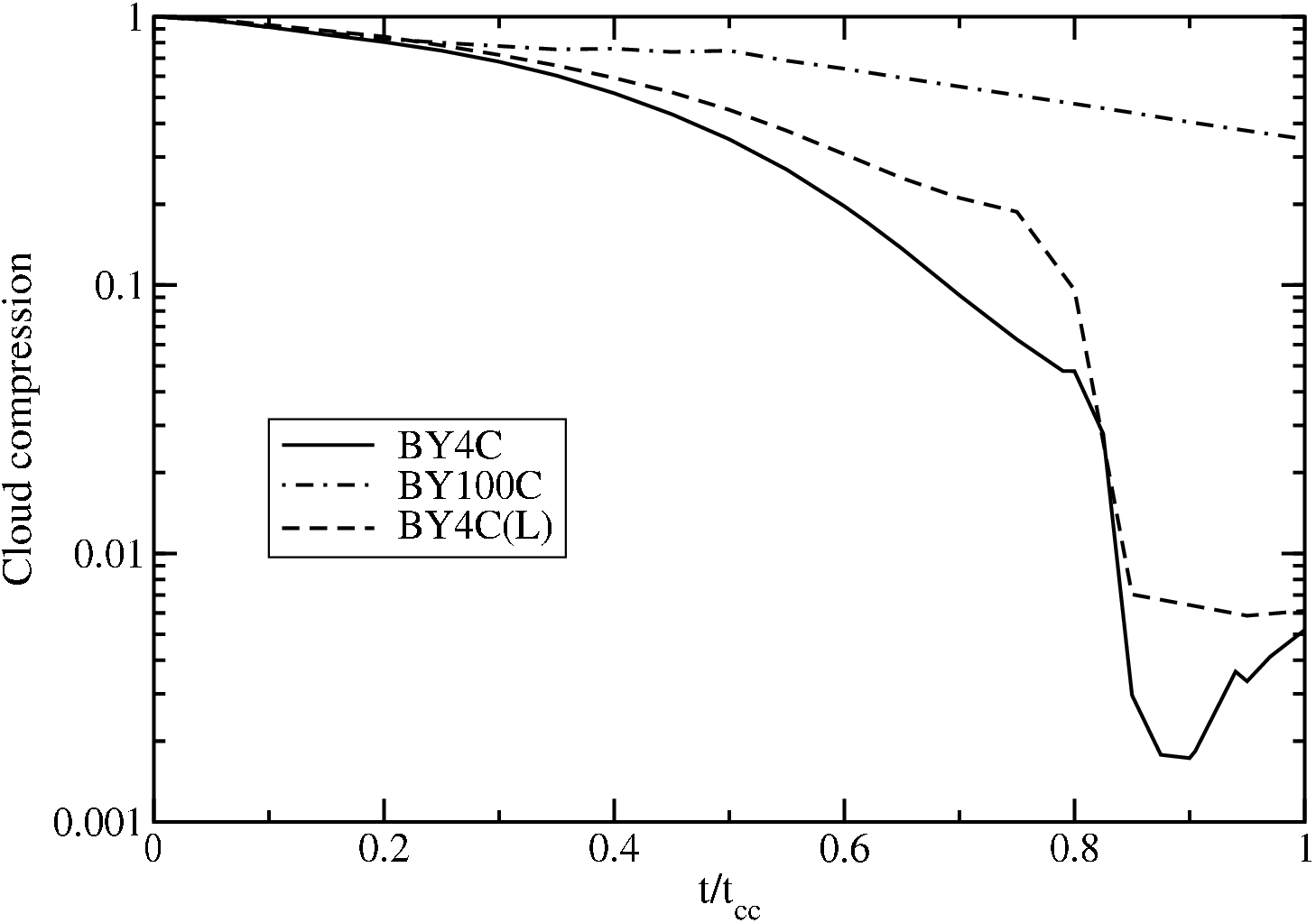}{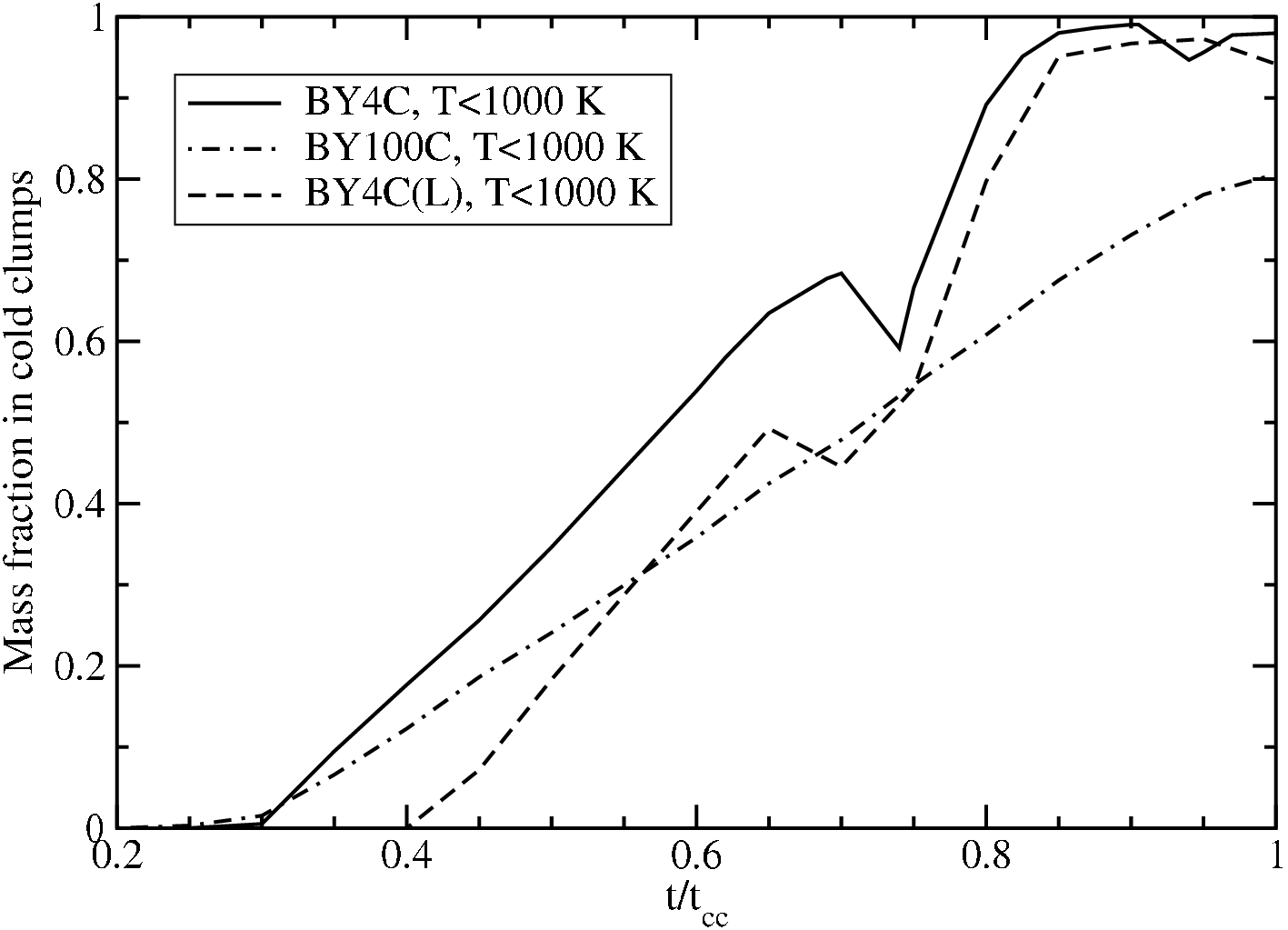}
\caption{Comparison of cloud compression ({\it a}) and cooling 
efficiency ({\it b}) for the radiatively-cooled $B_y$ runs.
}
\label{fig:by}
\end{figure}

For the $\beta_0=4$, $B_y$ field case (BY4C), 
we find that the increased compression 
from the trapped field lines greatly increases the cooling 
efficiency over the timescales considered.  One can then ask, how 
weak must such a field be in order to not dramatically enhance the 
compression and cooling over the same timescale?  
We therefore construct 
an additional $B_y$ run with $\beta_0=100$ (BY100C).  The cloud compression 
and cooling efficiencies for the different $B_y$ runs are presented 
in Figure \ref{fig:by}.  As we can see, the $\beta_0=100$ case 
behaves similarly to an unmagnetized cloud over the timescale 
considered.  Nevertheless, as the magnetic field lines continue 
to build up on the nose of the cloud, even this initially weak field 
will eventually play an important role in the 
cloud evolution, although over a longer timescale than the 
cloud compression time.

Finally, we find that the magnetic fields in the 
$B_x$ case with $\beta_0=4$ (BX4C) have 
little effect on the compression or cooling of the cloud.  It is 
worth considering whether a stronger initial field might change 
this conclusion.  Therefore we consider a $B_x$ run with 
$\beta_0=1$ (run BX1C).  In Figure \ref{fig:bx} we present the 
cloud compression and cooling efficiencies for the different $B_x$ 
runs.  We find that even a $B_x$ field in initial equipartition
with the thermal gas pressure has little influence on 
the compression or cooling of the cloud. A magnetic field component
aligned with the direction of shock propagation thus appears to
have little influence on the dynamic or thermal evolution
of radiative clouds.

\subsection{Effects of Numerical Resolution}
\label{sec:resolution}
Next we consider the effect of numerical resolution on our results.  
Run BY4C(L) uses an identical setup to run BY4C, 
but with half the resolution.  
Comparing these two runs gives us some idea of how well 
converged our solutions are.  The high compression and efficient 
cooling in the $B_y$ runs pose the most demanding 
resolution requirements of all the runs considered, so this example
represents a worst case comparison of convergence.

Figure \ref{fig:by} includes a comparison of the cloud compression and cooling 
efficiencies as a function of time for these two runs.  
Although there is a significant time 
lag in the cooling and compression for run BY4C(L), 
the asymptotic values for the cloud compression and mass fraction 
cooled to $<1000K$ generally 
agree well between the two runs ($\lesssim 10$\% differences).  
The time lag can be attributed to
the more diffusive nature of low resolution grids, which tend
to smear out concentrated density peaks and thereby increase
the cooling times. This is particularly troublesome for low
temperature coolants which generally require the resolution
of much smaller spatial scales to capture the transition
through the cooling plateau, and maintain their edge against
thermalization effects arising from numerical viscosity.

We have already noted that many of our radiatively cooled runs reach 
a resolution limit toward the end of our simulations.  This limit 
currently prevents us from reliably extending the 
duration of our simulations.  
Noting that the cold, dense cloud remnants occupy very few 
cells on the grid toward the ends of the simulations, it becomes 
clear that the most efficient approach to resolving the late-time 
evolution is to use an adaptive mesh scheme.  
This capability is currently being added to our code, and results 
will be presented in future work.

\subsection{Effects of Artificial Resistivity}
\label{sec:resistivity}
Finally we consider the role of reconnection in our results and 
compare the effectiveness of different numerical methods for dealing with 
it.  We expect reconnection to be important whenever oppositely 
directed field lines come into close proximity.  We note that this 
can not happen in our $B_z$ runs since the field lines are 
prevented from bending or otherwise becoming oppositely directed 
by the enforced symmetry of our two-dimensional Cartesian geometry.  
For our $B_x$ runs, reconnection is most 
likely to happen near the surface of the cloud, near the Mach 
stem downwind, or in any vortices that form.  However, because most of
these events are external to the cloud, we do not expect reconnection to 
play an important role in its evolution.  
Furthermore, we have already shown that $B_x$ fields, in general, 
have little effect on the evolution of our cloud models, 
so it seems unlikely that reconnection would be important to our conclusions 
in that case.  For our $B_y$ runs, reconnection will 
be most important near 
the current sheet that forms along the symmetry plane downwind of 
the cloud (see panels E and F of Figure \ref{fig:density}).  
When field lines on opposite sides of the current sheet 
are squeezed closer together than a zone width, numerical 
reconnection occurs and regions of the field 
transform into closed field loops, 
evident in panel F of Figure \ref{fig:density}.  
The current sheet and reconnection events are also evident as 
very high $\beta$ regions in panels E and F of Figure \ref{fig:beta}; 
$\beta$ is high in these regions since reconnection events are 
characterized by the conversion of magnetic energy into thermal energy.  

Clearly the way these reconnection events are handled 
numerically may have an effect on the final outcome of the simulations.  
As described in \S 
\ref{sec:methods}, Cosmos tracks reconnection through 
the use of a resistivity term designed to spread current 
sheets out so they are resolved and recapture any magnetic 
energy lost in reconnection through the internal energy equation.  
Up to this point all of the runs have used the 
\citet{sto01} form of resistivity $\eta_{SP}$ 
with $k_1=0.1$ (equation \ref{eq:etasp}).  
We now consider one run with $\eta=0$ [run BY4C(L1)] and another using 
the \citet{nit01} form $\eta_N$ with $k_1=0.5$ (equation \ref{eq:etan}) 
[run BY4C(L2)].  Since we expect our $B_y$ runs to be most affected 
by reconnection, we tailor these runs after 
run BY4C(L) from the previous section.  Other than the form of $\eta$, 
these runs are all identical.  We find that $B_{max}/B_i$, 
$\rho_{max}/\rho_{cl,i}$, and $T_{min}$ are very similar in all 
three runs (Table \ref{tab:results}), indicating that the form of 
artificial resistivity used has little impact on the cloud 
evolution during these runs.  Differences between these runs are 
more evident in the downwind flow where, for instance, 
$\beta_{min}$ varies by almost an order of magnitude.

\section{Discussion and Summary}
\label{sec:discussion}

We have presented results from a series of two-dimensional shock-cloud 
simulations with the goal of highlighting the importance of different physical
processes, including the interplay between hydrodynamic, radiative, and
magnetic effects. These simulations represent the first such
calculations we know of that simultaneously consider magnetic fields and 
radiative cooling.  To facilitate easy comparison, we included runs 
with different combinations of physical processes active.  
We summarize our main results as follows:

1.  Unmagnetized, non-radiative 
clouds are destroyed on a few dynamical timescales through hydrodynamic 
(Kelvin-Helmholtz and Rayleigh-Taylor) instabilities \citep{kle94}.

2.  In the cooling-dominated regime, radiative clouds are {\em not} 
destroyed.  Instead, they 
form dense, cold filaments, which are 
presumably the precursors to active star-forming regions \citep{mel02,fra04a}.

3.  Tension in magnetic field lines along the surface of a cloud 
can suppress the growth of hydrodynamic instabilities, thus 
increasing the cloud's survivability even without radiative cooling 
\citep{mac94,jon96}.  This is true for our external field cases 
($B_x$ and especially $B_y$) whenever the fields achieve 
sufficient strength ($\beta \lesssim 10^4$ to suppress Rayleigh-Taylor 
and $\beta \lesssim 0.02$ to suppress Kelvin-Helmholtz).  
On the other hand, internal fields that do not 
thread through the cloud surface 
(such as our $B_z$ cases) are unable to suppress the growth 
of hydrodynamic instabilities.

4.  External magnetic field lines that are stretched over the 
surface of a cloud can greatly enhance its compression.  For 
radiative clouds, this can dramatically enhance the cooling 
efficiency.  For instance, the fraction of cloud material cooling 
from $10^4$ K to below 100 K increases from $\sim 0.7$ without 
magnetic fields (run C) to $>0.9$ with a $\beta_0=4$, $B_y$ field 
(run BY4C).  This enhancement is negligible, however, 
for the $B_x$ field orientation or an initially weak field aligned
along $B_y$ ($\beta_0 > 100$).

5.  Internal magnetic field lines resist compression in a cloud.  
For radiative clouds, this can dramatically reduce the cooling 
efficiency.  For instance, in simulations with a $B_z$ field of only 
modest initial strength ($\beta_0\lesssim100$), the cloud material is 
prevented from cooling below 100 K.  A very strong initial field 
($\beta_0\sim1$) can even prevent cooling below 1000 K.

Obviously, the field configurations studied in this work are highly 
idealized.  Real astrophysical fields are likely to have much more 
complex topologies \citep{val03}.  In particular, magnetic fields in galactic 
star-forming clouds are probably linked to 
underlying turbulent flow patterns \citep{mac04}.  In that context, 
magnetic fields are important in stabilizing such clouds against collapse, 
a role similar to that of the {\em internal} fields in this work.  
However, those star-forming clouds are often very cold ($T<100$ K) 
and subject to important physical process not considered in this 
work, such as neutral-ion drift, so we caution the reader against 
attempting to extrapolate our results to that regime. 

Due to resolution requirements
and computational limitations, the simulations in this work 
were carried out in two-dimensional
Cartesian geometry, and so the clouds represent slices through infinite
cylinders.  This special geometry likely affects some of our 
conclusions.  Here we speculate on how these results 
may change in three-dimensional simulations:

Since the initial
compressive shock in the cloud is highly symmetric, 
the additional convergence expected
in three-dimensional models might lead to stronger compressions and
enhanced cooling in radiative clouds even without magnetic fields.  
Three-dimensional simulations would also provide
an additional degree of freedom for fragmentation through dynamical
instabilities.

The role of magnetic fields in three dimensions may be more complicated.  
For a spherical cloud, we lose the distinction between 
our $B_y$ and $B_z$ runs.  Such transverse fields will cause 
enhanced compression of the cloud along one direction (the initial direction 
of the field), but will 
be unable to prevent expansion in the perpendicular direction.  
This lateral 
expansion can enhance the growth rate of the Rayleigh-Taylor 
instability \citep{gre99,gre00}, an action that was prevented 
in this work by the assumed symmetry of the two-dimensional runs.  
Thus, in contrast to two-dimensional 
results, magnetic fields in three-dimensional simulations can
hasten the destruction of non-radiative clouds.
It remains to be seen how radiative cooling 
would modify this conclusion.

Self-gravity has been neglected in this study on account of the 
constrained cylindrical geometry. Although the cloud parameters
are specifically chosen such that self-gravity is negligible 
initially, the local free-fall timescale becomes significantly
shorter during the latter stages of compression.  
For the runs with radiative cooling active, self-gravity becomes important at
approximately the time we stop the simulations.  It would, therefore, 
be interesting 
to follow shock-cloud simulations in three-dimensions with 
radiative cooling, magnetic fields, and self-gravity included.  
To do this at comparable resolution to that used in this 
work will require the use of adaptive mesh
refinement to concentrate resolution around the cloud fragments.
Adaptive gridding is currently being added to the Cosmos code, 
and we plan to revisit this problem in future work.

\begin{acknowledgements}
The authors would like to thank the VisIt development team at 
Lawrence Livermore National Laboratory (http://www.llnl.gov/visit/), 
in particular Hank Childs and Akira Haddox, for visualization 
support.  This work was performed
under the auspices of the U.S. Department of Energy by
University of California, Lawrence
Livermore National Laboratory under Contract W-7405-Eng-48.
\end{acknowledgements}

\clearpage
\appendix
\section{MHD Code Verification}
Here we review some of the tests used to verify 
and validate the MHD coding of Cosmos.
We consider transverse magnetic field pulse advection, 
circularly polarized Alfv\'en waves, 
sheared Alfv\'en waves, 
gas and magnetosonic rarefaction waves,
MHD Riemann problems,
and shock-cloud collisions. 
We do not discuss in detail the setup of any
of these tests since they can all be found in the
literature, which we reference where appropriate.
The first test, 
the advection of transverse magnetic field pulses, yields field 
profiles identical to the van Leer results in
Figure 1b of \citet{sto92} and we do not discuss this problem 
further.  In the following paragraphs we summarize
the results from each of the remaining tests.

To test the ability of Cosmos in handling smooth flows and to
evaluate the convergence order of our methods, we consider 
the traveling circularly polarized Alfv\'en wave test from \citet{tot00}.  
We calculate the mean-relative error 
(defined as $\bar{\epsilon}_\mathrm{rel}^n(a)
= \sum_{i,j,k} \vert a_{i,j,k}^n - A_{i,j,k}^n \vert / \sum_{i,j,k} 
\vert A_{i,j,k}^n \vert$,
where $a_{i,j,k}^n$ and $A_{i,j,k}^n$ are the numerical and exact 
solutions, respectively) 
for $B_\perp$ and $v_\perp$ as a function of grid resolution ($n$).
Here $B_\perp = B_y \cos \alpha - B_x \sin \alpha$ is the magnetic 
field component perpendicular to the direction of wave propagation, 
which is at 
an angle $\alpha = 30^\circ$ relative to the $x$ axis.  The perpendicular 
velocity component $v_\perp$ is calculated similarly.  
For $B_\perp$ we find errors 
$\bar{\epsilon}_\mathrm{rel}^8(B_\perp) = 1.983$, 
$\bar{\epsilon}_\mathrm{rel}^{16}(B_\perp) = 0.599$, 
$\bar{\epsilon}_\mathrm{rel}^{32}(B_\perp) = 0.133$, and 
$\bar{\epsilon}_\mathrm{rel}^{64}(B_\perp) = 0.033$.  
For $v_\perp$ we find errors 
$\bar{\epsilon}_\mathrm{rel}^8(v_\perp) = 0.985$, 
$\bar{\epsilon}_\mathrm{rel}^{16}(v_\perp) = 0.369$, 
$\bar{\epsilon}_\mathrm{rel}^{32}(v_\perp) = 0.114$, and 
$\bar{\epsilon}_\mathrm{rel}^{64}(v_\perp) = 0.041$.  
The averages of 
our errors at each resolution are similar to the averages 
reported in \citet{tot00} for the Flux-CD/CT scheme.  The errors converge 
at approximately second order.

Figure 18 in \citet{sto92} clearly demonstrates the need to test
thoroughly the limit of sheared Alfv\'en wave propagation,
which can generate unacceptable levels of dispersive error.
This class of tests has led to the development of a more stable
method of characteristics to compute properly
centered electromotive and Lorentz forces. We adopt
a similar approach here and use the Alfv\'en characteristic
equation to estimate causal interpolants and predict
time-averaged estimates of the magnetic and velocity
fields used as sources in the transverse components
of equations (\ref{eqn:mom}) and (\ref{eqn:b}),
in particular the $({\bf B} \cdot \nabla) {\bf B}$
and $({\bf B} \cdot \nabla) {\bf v}$ terms.
Our results are identical to Figures 18 and 19 of \citet{sto92}
for the two cases in which we use conventional differencing
and the method of characteristics, respectively.

The ability of Cosmos to capture and propagate nonlinear waves and shocks 
is evaluated with the MHD analog of the classic Sod shock tube problem 
of hydrodynamics introduced by \citet{bri88}. Since there is no
known analytic solution for this problem, we test the convergence 
of our numerical solutions using a self-convergence test.  For two 
successive runs with a resolution ratio of two, we calculate the 
$L$-1 norm error (i.e.,
$L_1^n (a) = \sum_{i,j,k} \Delta x_i \Delta y_j \Delta z_k
\vert a_{i,j,k}^n - a_{i,j,k}^{2n} \vert$, where
$a_{i,j,k}^n$ and $a_{i,j,k}^{2n}$ are the numerical 
solutions for $n$ and $2n$ zones, respectively, and
$j=k=\Delta y_j = \Delta z_k =1$ for this 1D problem).  For the fluid density
we find errors $L_1^{64} (\rho) = 0.00691$, $L_1^{128} (\rho) = 0.00362$, 
$L_1^{256} (\rho) = 0.00211$, and $L_1^{512} (\rho) = 0.00127$.  
As expected for shock problems, the convergence is approximately first order,
and our results are similar to \citet{sto92}.

For certain one-dimensional problems, \citet{fal02} reported significant 
errors when using the publically available ZEUS MHD code.  
Since Cosmos, as applied in this work, 
uses a ZEUS-like MHD scheme, we give particular emphasis 
to investigating these problems.  Our results show that Cosmos performs 
significantly better than the version of ZEUS used in \citet{fal02},
due in part to the introduction of artificial resistivity.
To illustrate this, in Figures 
\ref{fig:rarefaction} and \ref{fig:riemann}, we 
show the Cosmos results when applied to the fast rarefaction 
and Riemann test problems from Figures 2 
and 6 of \citet{fal02}.  In each figure, we include results 
with both the \citet{sto01} and 
\citet{nit01} forms of artificial resistivity 
and with no artificial resistivity ($\eta=0$).  The resistivity 
coefficient is set to $k_1=0.5$ for the fast rarefaction and 
$k_1=1.0$ for the Riemann problem.  In both problems, the Courant factor 
is set to 0.5, and the linear and quadratic artificial viscosity 
coefficients are set to 0.2 and 2.0, respectively.  For the fast 
rarefaction, Cosmos performs much better than the ZEUS results
shown in \citet{fal02}. Furthermore, we note that Cosmos 
is also less diffusive than the upwind method employed in \citet{fal02} 
when either no resistivity or the \citet{sto01} form of resistivity are used.
When the \citet{nit01} form is used, our results appear similar to the 
upwind method.  
Cosmos also significantly outperforms the ZEUS code on the Riemann problem 
and successfully captures all of the features, due predominately to the
addition of artificial resistivity.
On this problem, the \citet{nit01} form of resistivity 
yields better results.  Next, we consider the smooth gas rarefaction and 
stationary shock problems 
from Figures 3 and 5 of \citet{fal02}.  For the smooth gas rarefaction, 
\citet{fal02} noted that, despite being a second-order code, ZEUS
yields only first-order convergence.  
For Cosmos, we find errors of 
$\bar{\epsilon}_\mathrm{rel}^{50} (\rho)=1.4 \times 10^{-2}$, 
$\bar{\epsilon}_\mathrm{rel}^{100} (\rho)=3.8 \times 10^{-3}$, 
$\bar{\epsilon}_\mathrm{rel}^{200} (\rho)=9.2 \times 10^{-4}$, 
$\bar{\epsilon}_\mathrm{rel}^{400} (\rho)=2.7 \times 10^{-4}$, and 
$\bar{\epsilon}_\mathrm{rel}^{800} (\rho)=8.4 \times 10^{-5}$, 
which converge at approximately second order, as expected.
On the stationary shock problem, 
although Cosmos does not achieve the same level
of accuracy as the upwind code
employed in \citet{fal02}, the error in Cosmos as measured from the 
analytical post-shock gas pressure plateau is about three times smaller
than for ZEUS. This, again, is attributed mostly to the addition 
of artificial resistivity.

For our final test we simulate the magnetized shock-cloud collision 
first described in \citet{dai98} and repeated in \citet{tot00}.  This 
test is particularly appropriate for the investigations considered 
in this work.  We investigate this problem 
using an $n \times n$ grid with $n=50$, 100, 200, and 400.  The 
mean-relative errors in density and $B^2$ as a 
function of resolution are 
$\bar{\epsilon}_\mathrm{rel}^{50} (\rho)=0.103$, 
$\bar{\epsilon}_\mathrm{rel}^{50} (B^2)=0.283$, 
$\bar{\epsilon}_\mathrm{rel}^{100} (\rho)=0.061$, 
$\bar{\epsilon}_\mathrm{rel}^{100} (B^2)=0.186$, 
$\bar{\epsilon}_\mathrm{rel}^{200} (\rho)=0.030$, and 
$\bar{\epsilon}_\mathrm{rel}^{200} (B^2)=0.086$, 
where we have used the high-resolution 
($400 \times 400$ zone) simulation as the reference solution.  Again 
the convergence is approximately first order as expected for this 
type of problem, and our errors are consistent with those reported 
by \citet{tot00}.

\clearpage
\bibliographystyle{apj}
\bibliography{myrefs}

\clearpage

\clearpage




\clearpage

\clearpage

\clearpage

\clearpage

\clearpage

\clearpage
\begin{figure}
\plottwo{f7a.eps}{f7b.eps}
\caption{Comparison of cloud compression ({\it a}) and cooling 
efficiency ({\it b}) for the radiatively-cooled $B_x$ runs.
}
\label{fig:bx}
\end{figure}

\clearpage
\begin{figure}
\plotone{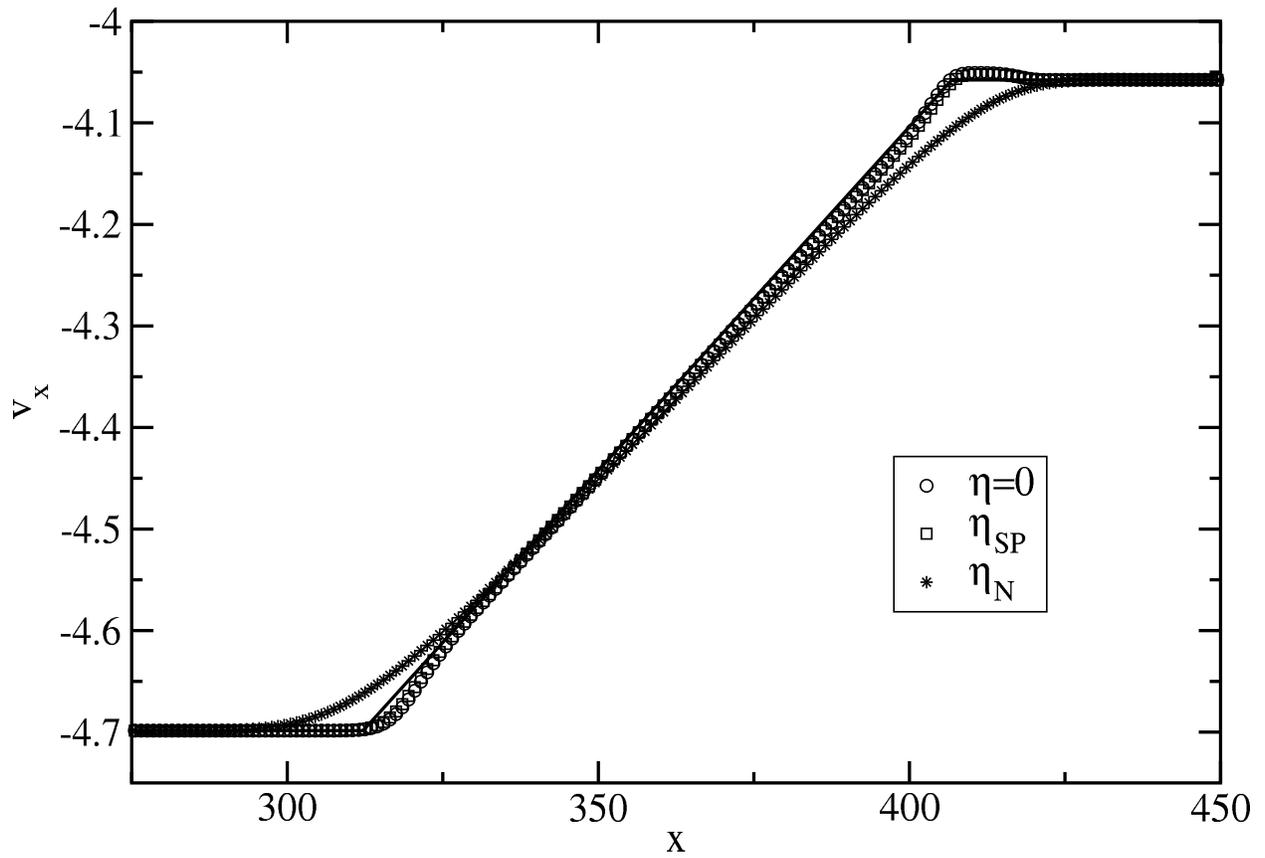}
\caption{Test of the fast rarefaction wave from Figure 2 of 
\citet{fal02}, comparing results using different forms of artificial 
resistivity.  The solid line gives the exact solution for this problem.
}
\label{fig:rarefaction}
\end{figure}

\clearpage
\begin{figure}
\plotone{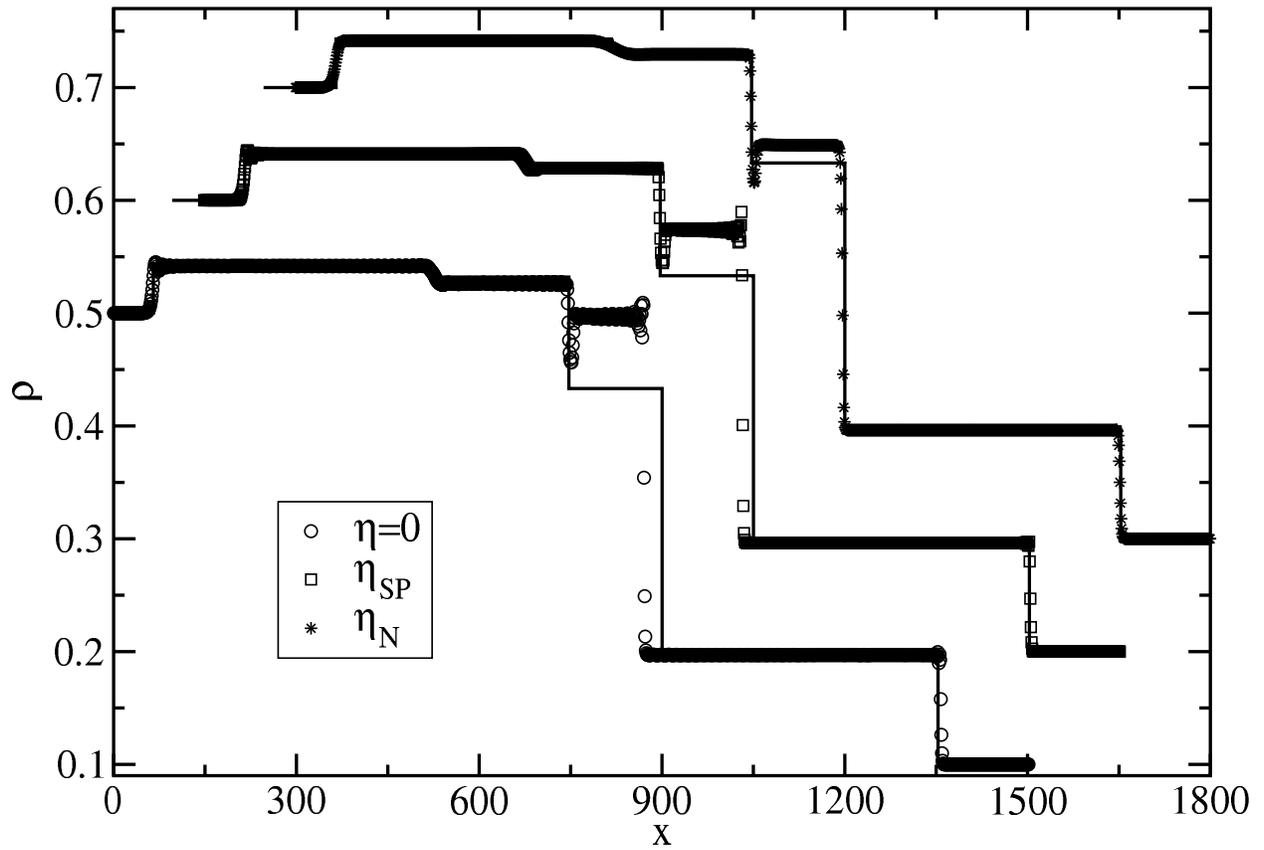}
\caption{Test of the Riemann problem from Figure 6 of \citet{fal02}, 
comparing results using different forms of artificial resistivity.  
The solid lines give the exact solution for this problem.  The results 
are offset for clarity.
}
\label{fig:riemann}
\end{figure}

\clearpage
\begin{deluxetable}{ccccc}
\tablewidth{0pt}
\tablecaption{Model Parameters \label{tab:models}}
\tablehead{
\colhead{Run} &
\colhead{Resolution\tablenotemark{a}} &
\colhead{$\beta_0$} &
\colhead{Field Component} & 
\colhead{Cooling}
}
\startdata
A       & 200 & $\infty$ &       & no  \\  
C       & 200 & $\infty$ &       & yes \\  
BZ4A    & 200 & 4        & $B_z$ & no  \\  
BZ4C    & 200 & 4        & $B_z$ & yes \\  
BY4A    & 200 & 4        & $B_y$ & no  \\  
BY4C    & 200 & 4        & $B_y$ & yes \\  
BX4A    & 200 & 4        & $B_x$ & no  \\  
BX4C    & 200 & 4        & $B_x$ & yes \\  
BZ1C    & 200 & 1        & $B_z$ & yes \\  
BZ100C  & 200 & 100      & $B_z$ & yes \\  
BY100C  & 200 & 100      & $B_y$ & yes \\  
BX1C    & 200 & 1        & $B_x$ & yes \\  
BY4C(L) & 100 & 4        & $B_y$ & yes \\  
BY4C(L1) & 100 & 4       & $B_y$ & yes \\  
BY4C(L2) & 100 & 4       & $B_y$ & yes     
\enddata
\tablenotetext{a}{Initial number of zones per cloud radius.}
\end{deluxetable}

\clearpage
\begin{deluxetable}{ccccc}
\tablewidth{0pt}
\tablecaption{Summary of Key Results at $t=t_{cc}$ \label{tab:results}}
\tablehead{
\colhead{Run} &
\colhead{$\beta_{min}$} &
\colhead{$B_{max}/B_i$} &
\colhead{$\rho_{max}/\rho_{cl,i}$} &
\colhead{$T_{min}$ (K)}
}
\startdata
A       & $\cdots$ & $\cdots$ & 1.5E1 & 5.3E4  \\  
C       & $\cdots$ & $\cdots$ & 8.7E2 & 2.4E1  \\  
BZ4A    & 4.1      & 1.3E1    & 1.3E1 & 4.4E4  \\  
BZ4C    & 1.9E-3   & 7.3E1    & 4.7E1 & 3.8E2  \\  
BY4A    & 1.6E-5   & 1.8E2    & 4.7E1 & 2.8E5  \\  
BY4C    & 3.8E-5   & 8.2E2    & 2.1E4 & 1.5E1  \\  
BX4A    & 2.9E-1   & 2.5E1    & 1.1E1 & 6.1E4  \\  
BX4C    & 4.0E-3   & 3.9E1    & 7.2E2 & 2.6E1  \\  
BZ1C    & 7.1E-4   & 4.2E1    & 4.2E1 & 3.0E2  \\  
BZ100C  & 1.1E-3   & 3.4E2    & 1.3E2 & 1.0E2  \\  
BY100C  & 5.4E-6   & 5.7E2    & 1.5E3 & 2.2E1  \\  
BX1C    & 3.7E-3   & 1.9E1    & 1.0E3 & 2.3E1  \\  
BY4C(L) & 2.8E-5   & 4.7E2    & 5.4E3 & 1.8E1  \\  
BY4C(L1) & 6.9E-5  & 4.8E2    & 4.4E3 & 1.9E1  \\  
BY4C(L2) & 1.1E-4  & 4.3E2    & 5.9E3 & 2.2E1      
\enddata
\end{deluxetable}


\end{document}